# Topological Hyperbolic and Dirac Plasmons

**Nahid Talebi**

Max Planck Institute for Solid State Research, Heisenbergstr. 1, 70569 Stuttgart, Germany

**Abstract** In this chapter, criteria for existence of propagating optical modes which are transversely bound at the interface of two materials are studied. In particular, quite general cases are considered, where the materials involved are assumed to be anisotropic, but also demonstrating magneto-electric effects. Moreover, surface states of two-dimensional materials like topological insulators and graphene are also modeled via consideration of a conductivity sheet existing at the interface. A characteristic equation for obtaining the propagation constant of generalized interface modes is presented. Furthermore, optical modes sustained by a thin film of anisotropic materials with magneto-electric effect and topological surface states are also investigated. It is shown that interface modes supported by such a system are hybrid in nature, and can be further decomposed into the well-known classes of transverse magnetic and electric modes, only at the absence of magneto-electric effect. Although the formulations driven here are mathematically abstract, they can be used to investigate polaritons in van der Waal materials, hyperbolic materials, and topological insulators.

## 1.1 Introduction

Charge density waves associated with a free-electron gas inside metals can couple to the electromagnetic waves at the surface of the metal and create surface plasmon polaritons (SPPs). SPPs are guided waves which are transversely bound to the surface of a metal and transport the electromagnetic energy. In other words, a surface of a metal is topologically the simplest considerable geometry for supporting guided waves. In addition SPPs have other fascinating characteristics, such as transporting [1-3] and trapping [4] of the electromagnetic energy beyond the diffraction limit, as well as enhanced light–matter interaction [5, 6] happening due to the ability of SPPs to localize light waves at the nanoscale [7, 8]. This localization is concomitant with quantization of either linear or angular momenta of the optical waves, in the Fabry-Pérot-like linear resonators like nanorods [9] and slits [10], or geometries supporting rotating modes [11, 12], respectively. More sophisticated localization mechanisms happen in geometries like tapers, which support both linear and angular momentum orders, and facilitate highly efficient near- to far-field coupling of optical resonances [13, 14]. A prominent characteristic of SPPs is however a large attenuation constant, which hinders them from suitable applications in technologically relevant fields dealing with the transport of information. Circumventing the problem of loss in metals has initiated the field of dielectric-based nanophotonics [15].

Concerning the electrodynamics, optical modes bound at interfaces may happen for other classes of materials besides metals. A well-known example of such modes is the Dyakonov wave which occurs at the surface of anisotropic dielectrics [16-18], or the Dyakonov plasmon when considering hyperbolic materials [19, 20], also called hyperbolic plasmons here. A hyperbolic material has a uniaxial crystalline structure with two distinguished permittivity components, namely in-plane ($\varepsilon_{xx} = \varepsilon_{yy} = \varepsilon_{\|}$) and normal ($\varepsilon_{zz} = \varepsilon_{\perp}$) components, where the coordinate system is positioned along the principal axes of the crystal. In addition, in some frequency ranges, the signs of the two permittivity components are not the same. In other words, for some certain polarizations of the incident light the material behaves like a metal, whereas for other polarizations the optical response signifies a dielectric-like behavior. The dispersion relation for a plane wave propagating in the bulk of a hyperbolic material at an arbitrary direction with the wave vector $\vec{k} = \left(k_x, k_y, k_z\right)$ is decomposed into two groups, for ordinary and extraordinary rays correspondingly. The isofrequency surface of the extraordinary rays (transverse magnetic waves) in particular is given by [21]

$$\frac{k_x^2 + k_y^2}{\varepsilon_{r\perp}(\omega)} + \frac{k_z^2}{\varepsilon_{r\|}(\omega)} = \frac{\omega^2}{c^2} \tag{1.1}$$

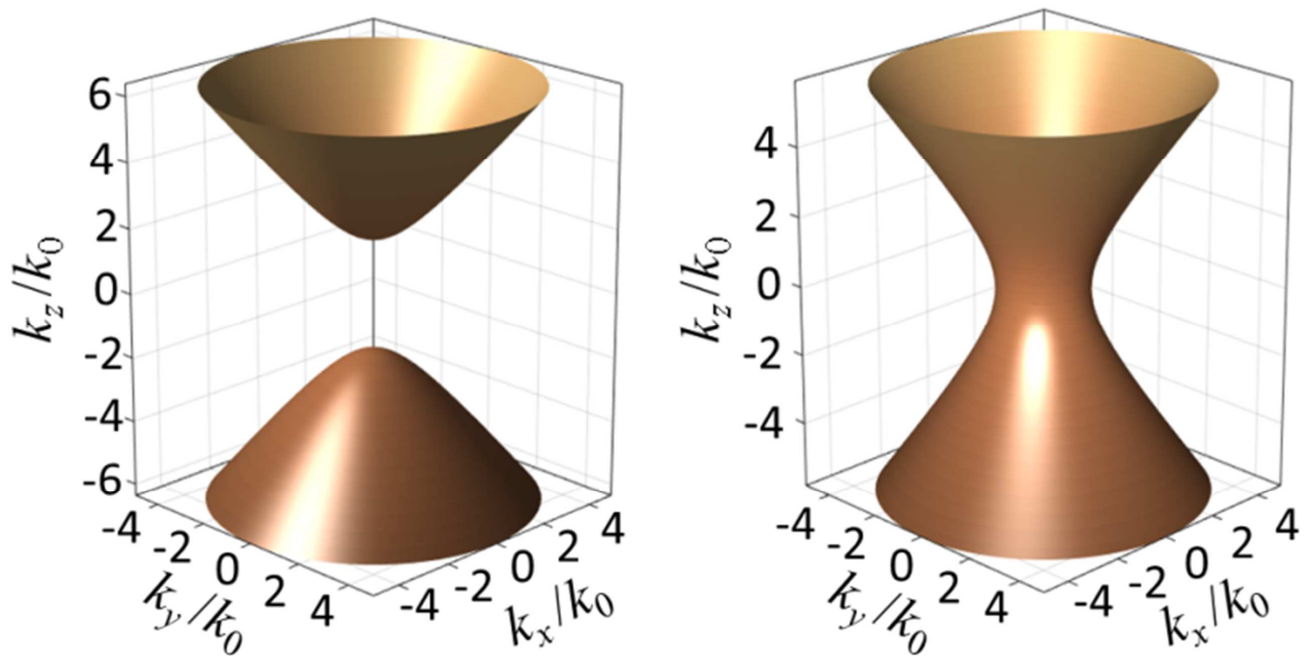

**Fig. 1.** Isofrequency surfaces of plane waves propagating inside a hyperbolic medium with (left) $\varepsilon_{r\|} < 0$ and $\varepsilon_{r\perp} > 0$, and (right) $\varepsilon_{r\|} > 0$ and $\varepsilon_{r\perp} < 0$.

which for the case of $\varepsilon_{r\perp}\varepsilon_{r\parallel} < 0$ forms a hyperboloid (Fig. 1). Additionally, two distinguished types of hyperboloids are expected: for the material with $\varepsilon_{r\parallel} < 0$ and $\varepsilon_{r\perp} > 0$ the isofrequency surface exhibits a gap and this material is referred to as hyperbolic type I, whereas for material with $\varepsilon_{r\parallel} > 0$ and $\varepsilon_{r\perp} < 0$ there is not any gap in the isofrequency surface. Such a material is called hyperbolic type II.

Why are hyperbolic materials interesting? Besides being a simpler case of a metamaterial [22], there exist several applications for hyperbolic materials. They can be used for enhancement of the Purcell factor and spontaneous emission [23], enhancing the photonic density of states (PDOS) [20], and they have an extreme confinement factor for the optical energy at the nano scale because of the large effective refractive indices of the optical modes sustained by hyperbolic materials [24]. Moreover, there are natural materials with hyperbolic dispersion covering distinct regions of the electromagnetic spectrum, from terahertz to the ultraviolet [25] (Fig. 2).

Interestingly, tetradymites and more specifically topological insulators (TIs) are naturally hyperbolic as well. Among natural hyperbolic materials bismuth-based TIs have been more intensively investigated considering their optical density of states [26-28]. The optical modes at the surface of TIs have contributions from Dyakonov plasmons, as well as electronic surface states (SSs). Very similar to graphene, two-dimensional plasmons also exist at the surface of TIs and sustain ultrahigh wavenumbers which allow for the confinement of modes up to three orders of magnitude smaller than the diffraction limit. Thanks to their gapless dispersion, they can cover an extremely large bandwidth for the energies $\hbar\omega > \sqrt{\hbar^2\omega_0^2 - \frac{1}{2}n_s\eta_0^2 e^4 v_F^2}$ , where a Drude model is considered for the local response conductivity [29]. Here $\omega_0$ is the damping rate, $v_F$ is the Fermi velocity, $e$ is the elementary charge, $h$ is the Planck constant, $\eta_0 = \sqrt{\mu_0/\varepsilon_0}$ is the free-space impedance, and $n_s$ is the carrier density. However, dissimilar to graphene, it is not easy to assess the role of SS versus the bulk contributions to the interface optical modes, when a pure imaging technique like scanning near-field electron microscopy (SNOM) is employed [30]. In contrast, methods like electron energy-loss spectroscopy which allows for momentum-resolved investigations can be used for directly resolving the dispersion [31].

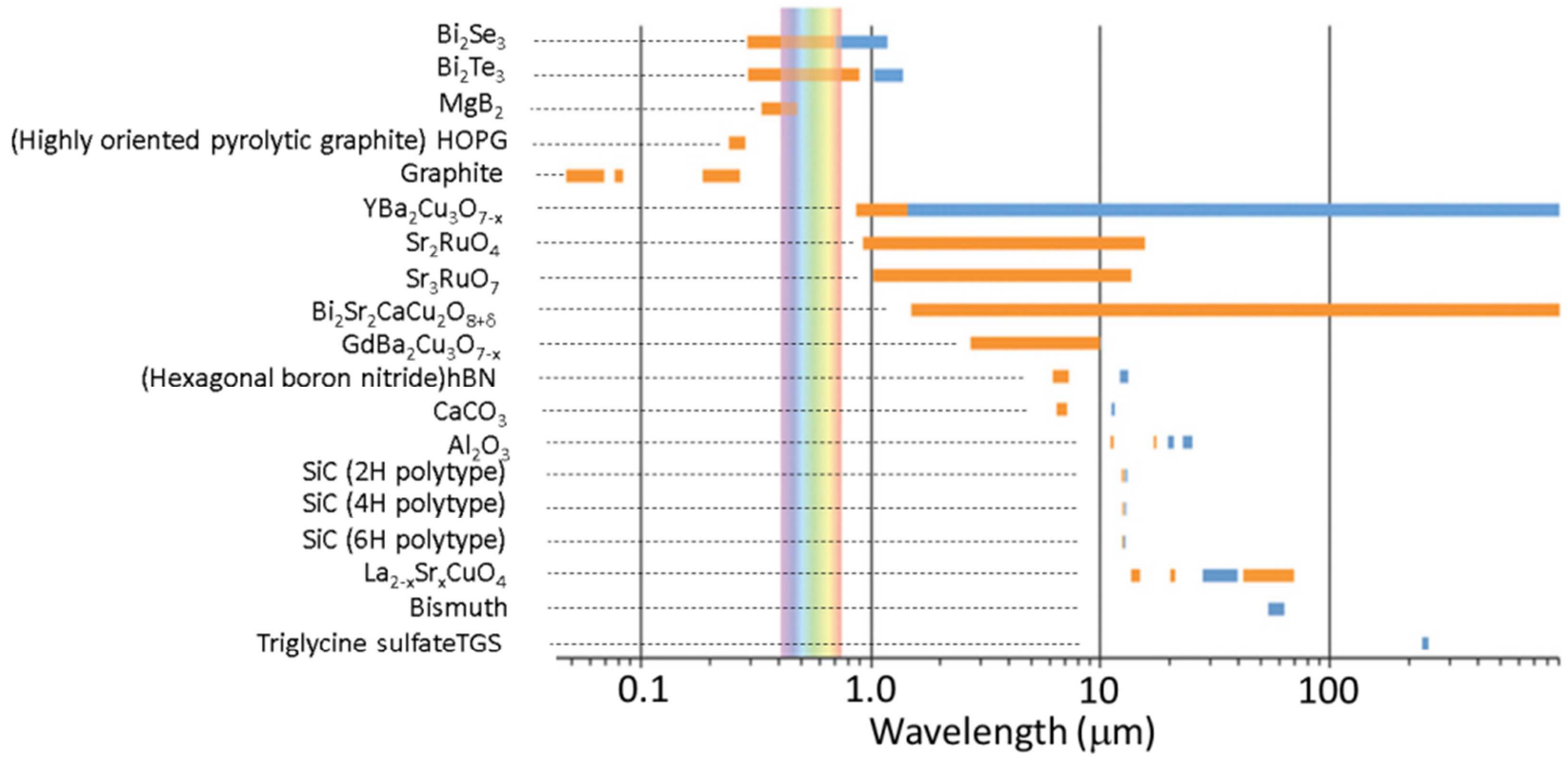

**Fig. 2.** Natural hyperbolic materials and the frequency range they cover. Adapted from ref. [25]. Blue and orange colours correspond to the bands with one and two negative components in the diagonal dielectric permittivity tensor, respectively.

In addition to the Dyakonov and Dirac plasmons, there is another contribution to be added to the surface optical modes, caused by the topological magneto-electric (ME) effect [32]. It has been shown recently that topological insulators are platform for the realization of axion electrodynamics [33-35], when the time reversal symmetry is weakly broken, for example by applying a static magnetic field. Axionic behavior in TIs originate from the pseudoscalar term of the ME response, and appears as an additional contribution to the free energy of the material as $W_{TI}\left(\vec{E},\vec{H}\right) \propto -\vec{E}\cdot\vec{H}$ . The ME response results also in the generalization of the constitutive relations of the Maxwell equations in the form given by:

$$\vec{D}(\vec{r},\omega) = \varepsilon_0\hat{\varepsilon}_r(\omega) : \vec{E}(\vec{r},\omega) + \varsigma\vec{B}(\vec{r},\omega) \tag{1.2a}$$

and

$$\vec{H}(\vec{r},\omega) = \frac{1}{\mu_0\mu_r(\omega)}\vec{B}(\vec{r},\omega) - \xi\vec{E}(\vec{r},\omega) \tag{1.2b}$$

where for the topological magneto-electric effect $\varsigma = \xi = \alpha\theta/\eta_0\pi$ [36, 37], where $\alpha = e^2/4\pi\varepsilon_0\hbar c$ is the fine-structure constant, $\theta$ is a phenomenological parameter in the effective Ginzburg–Landau theory describing the topological ME effect, and $\theta = \pi$ . $\hat{\varepsilon}_r$ is the permittivity tensor and $\mu_r$ is the permeability, $\varepsilon_0$ and $\mu_0$ are the free-space permittivity and permeability, $\vec{E}$ and $\vec{H}$ are the electric and magnetic field components respectively, $\vec{D}$ is the displacement vector, and $\vec{B}$ is the magnetic flux density. We further mention here, that the topological ME effect can also be modelled by the off-diagonal element of the two-dimensional conductivity tensor at the interface ($\sigma_{xy}$ and $\sigma_{yx}$ elements). However, to distinguish between the Dirac plasmon and the topological ME effect, and to further include the chiral ME effect, we chose to use eqs. (1.2a) and (1.2b).
To fully understand the behavior of the optical modes at the interface of TIs with other materials, all the important contributions stated above should be taken into account: (i) the hyperbolic nature of the material, (ii) SSs and Dirac plasmons associated with them, and (iii) the topological ME effect. In this chapter a simple model will be derived and proposed to investigate the full PDOS at the surface of TIs, which is also applicable to heterostructures like graphene upon a hyperbolic material like hexagonal boron nitride (hBN). The aim is to maintain the discussions at a mathematically abstract level to be able to provide the reader with the general idea. However, few examples will be provided for physically relevant systems like $Bi_2Se_3$, graphene, and hBN.

## 1.1 Helmholtz Theory for Hyperbolic Materials with ME effect

For isotropic materials without ME effect, it is straightforward to derive the Helmholtz equation for the electric and magnetic field components individually in the form of $\nabla^2\vec{E}+\varepsilon_r(\omega)\mu_r(\omega)k_0^2\vec{E}=0$ and $\nabla^2\vec{H}+\varepsilon_r(\omega)\mu_r(\omega)k_0^2\vec{H}=0$ , where a source-free medium has been considered. Here $k_0^2=\omega^2\varepsilon_0\mu_0$ is the free space wavenumber. However, for reducing the number of equations and the possibility to decompose the modes into symmetry groups like transverse electric and transverse magnetic waves, it is better to use potentials. Moreover, for anisotropic materials with ME effect, it is not possible to propose wave equations for field components. The Helmholtz equation is however derivable for some specific potentials, such as the magnetic vector potential ($\vec{A}$) and the scalar potential ($\varphi$) pairs [38]. Such a method demands a generalization of the Lorentz gauge theory, as will be shown here.

By using the relation between the potentials and field components as $\vec{B}=\vec{\nabla}\times\vec{A}$ and $\vec{E}=-i\omega\vec{A}-\vec{\nabla}\varphi$, Maxwell's equations, and the constitutive relation given by eqs. (1.2a) and (1.2b), one can derive the following equation for the magnetic vector potential

$$\begin{aligned}\frac{1}{\mu_0\mu_r}\left(\vec{\nabla}\vec{\nabla}\cdot\vec{A}-\nabla^2\vec{A}\right)-\omega^2\varepsilon_0\hat{\varepsilon}_r:\vec{A}+\\ i\omega\varepsilon_0\hat{\varepsilon}_r:\vec{\nabla}\varphi+i\omega\left(\xi-\varsigma\right)\vec{\nabla}\times\vec{A}=0\end{aligned} \tag{1.3}$$

A gauge theory can be used to derive the Helmholtz equation for $\vec{A}$:

$$\vec{\nabla}\varphi = \frac{-\hat{\varepsilon}_r^{-1}}{i\omega\varepsilon_0\mu_0\mu_r} : \vec{\nabla}\vec{\nabla}\cdot\vec{A} - \left(\xi - \varsigma\right)\frac{\hat{\varepsilon}_r^{-1}}{\varepsilon_0} : \vec{\nabla}\times\vec{A} \tag{1.4}$$

which is referred to as the generalized Lorentz gauge. Using (1.3) and (1.4), we derive the Helmholtz equation for the magnetic vector potential as $\nabla^2\vec{A} + k_0^2\mu_r(\omega)\hat{\varepsilon}_r(\omega) : \vec{A} = 0$. Note that eq. (1.4) does not allow for a direct relation between $\varphi$ and $\vec{\nabla}\cdot\vec{A}$, as usual. However, what we need for deriving the field components is $\vec{\nabla}\varphi$ and not $\varphi$ itself. Moreover, for the topological ME effect where $\xi = \varsigma$, eq. (1.4) will be simplified to the Lorentz gauge. Finally, the field components are given by

$$\vec{E} = -i\omega\vec{A} + \frac{\hat{\varepsilon}_r^{-1}}{i\omega\varepsilon_0\mu_0\mu_r} : \vec{\nabla}\vec{\nabla}\cdot\vec{A} - \left(\xi - \varsigma\right)\frac{\hat{\varepsilon}_r^{-1}}{\varepsilon_0} : \vec{\nabla}\times\vec{A} \tag{1.5}$$

for the electric field, and

$$\vec{H} = \frac{1}{\mu_0\mu_r}\vec{\nabla}\times\vec{A} - \xi\vec{E} \tag{1.6}$$

for the magnetic field.

## 1.2 Optical modes at a single interface

We consider now an interface between two hyperbolic materials which sustain topological ME effect. The interface is located at $z = 0$, with the only nonzero permittivity components $\varepsilon_{r\,xx} = \varepsilon_{r\,yy} = \varepsilon_{r\parallel}$ and $\varepsilon_{r\,zz} = \varepsilon_{r\perp}$. We allow also for surface states modelled by a two dimensional conductivity $\sigma\left(\omega, k_\perp\right)$, which only affects the boundary conditions in the formulations as $\hat{z}\times\left(\vec{H}_1 - \vec{H}_2\right) = \sigma\vec{E}$, where $\vec{H}_1$ and $\vec{H}_2$ are magnetic field components at $z \geq 0$ and $z \leq 0$, respectively. Without loss of generality, we assume that the optical waves propagate along the *x* axis, are invariant along the *y*-axis, and are evanescent with respect to the *z*-axis. The solutions for the vector potential can be then constructed as

$$A_\alpha\left(\vec{r}, \omega\right) = \tilde{A}_1^\alpha \exp\left(-\kappa_z^{(1,\alpha)} z\right)\exp\left(-i\beta x\right) \tag{1.7a}$$

for the region $z \geq 0$, and

$$A_\alpha\left(\vec{r},\omega\right) = \tilde{A}_2^\alpha \exp\left(+\kappa_z^{(2,\alpha)} z\right)\exp\left(-i\beta x\right) \tag{1.7b}$$

for the region $z \leq 0$. $\beta = \beta' - i\beta''$ is the complex propagation constant, $\left(\kappa_z^{(j,\alpha)}\right)^2 = \beta^2 - \varepsilon_{r\alpha\alpha}^j \mu_r^j k_0^2$, $\alpha \in \left(x,y,z\right)$, and $j = 1,2$ designates the domains $z \geq 0$ and $z \leq 0$, respectively. $\tilde{A}_1^\alpha$ and $\tilde{A}_2^\alpha$ are unknown coefficients for the vector potential expansions in each domain.

The boundary conditions should be satisfied at $z = 0$ for the tangential electric and magnetic field components. Obviously there are only 4 equations from which to obtain the unknowns $\left(\tilde{A}_1^x, \tilde{A}_2^x\right)$, $\left(\tilde{A}_1^y, \tilde{A}_2^y\right)$, and also $\left(\tilde{A}_1^z, \tilde{A}_2^z\right)$. In order to avoid an underdetermined system of equations, pairs in the form of $\left(A_x, A_y\right)$, $\left(A_y, A_z\right)$, and $\left(A_x, A_z\right)$ should be used. After some straightforward algebra, it is furthermore understood that pure TM$_x$, TM$_y$, or TM$_z$ modes, as well as $\left(A_x, A_z\right)$ modes will not satisfy the boundary conditions. The modes associated to the $\left(A_x, A_y\right)$ and $\left(A_y, A_z\right)$ pairs are denoted here by $A^{xy}$ and $A^{yz}$, respectively.

After meeting all the requirements stated above, the characteristic equation for the $A^{xy}$ propagating modes is obtained as:

$$\begin{aligned}
&\left(\frac{\kappa_z^{(1,y)}}{\mu_{r1}} + \frac{\kappa_z^{(2,y)}}{\mu_{r2}} + i\omega\mu_0\sigma\left(\omega,\beta\right)\right)\times \\
&\left(\frac{\kappa_z^{(1,x)}}{\varepsilon_{r\|}^1} + \frac{\kappa_z^{(2,x)}}{\varepsilon_{r\|}^2} + \frac{\sigma\left(\omega,\beta\right)}{i\omega\varepsilon_0}\frac{\kappa_z^{(1,x)}}{\varepsilon_{r\|}^1}\frac{\kappa_z^{(2,x)}}{\varepsilon_{r\|}^2}\right) \\
&= -\eta_0^2 \frac{\kappa_z^{(1,x)}}{\varepsilon_{r\|}^1}\frac{\kappa_z^{(2,x)}}{\varepsilon_{r\|}^2}\left(\xi^1 - \xi^2\right)^2
\end{aligned} \tag{1.8}$$

where $\xi^1$ and $\xi^2$ are the topological ME indices for domains $z \geq 0$ and $z \leq 0$, respectively. For the $A^{yz}$ modes, the characteristic equation is obtained as:

$$\left(\frac{\kappa_z^{(1,y)}}{\mu_{r1}}+\frac{\kappa_z^{(2,y)}}{\mu_{r2}}+i\omega\mu_0\sigma\left(\omega,\beta\right)\right)\times$$
$$\left(\frac{\kappa_z^{(1,z)}}{\varepsilon_{r\|}^1}+\frac{\kappa_z^{(2,z)}}{\varepsilon_{r\|}^2}+\frac{\sigma\left(\omega,\beta\right)}{i\omega\varepsilon_0}\frac{\kappa_z^{(1,z)}}{\varepsilon_{r\|}^1}\frac{\kappa_z^{(2,z)}}{\varepsilon_{r\|}^2}\right) \tag{1.9}$$
$$=-\eta_0^2\frac{\kappa_z^{(1,z)}}{\varepsilon_{r\|}^1}\frac{\kappa_z^{(2,z)}}{\varepsilon_{r\|}^2}\left(\xi^1-\xi^2\right)^2$$

The only difference between the characteristic equations (1.8) and (1.9) is the replacement of $\kappa_z^{(x,2)}$ by $\kappa_z^{(z,2)}$ in the second term on the left side and also on the right side. Moreover, both equations are similar when $\varepsilon_{r\|}^i=\varepsilon_{r\perp}^i$, i.e. when the material is isotropic. In other words, $A^{xy}$ and $A^{yz}$ modes become degenerate for an isotropic material.

For materials without a topological ME effect, eqs. (1.8) and (1.9) form three individual groups of modes, which are $\left(\kappa_z^{(1,x)}/\varepsilon_{r\|}^1\right)+\left(\kappa_z^{(2,x)}/\varepsilon_{r\|}^2\right)+\sigma\left(\omega,\beta\right)\kappa_z^{(1,x)}\kappa_z^{(2,x)}/i\omega\varepsilon_0\varepsilon_{r\|}^1\varepsilon_{r\|}^2=0$, $\kappa_z^{(1,y)}/\mu_{r1}+\kappa_z^{(2,y)}/\mu_{r2}+i\omega\mu_0\sigma\left(\omega,\beta\right)=0$, and $\left(\kappa_z^{(1,z)}/\varepsilon_{r\|}^1\right)+\left(\kappa_z^{(2,z)}/\varepsilon_{r\|}^2\right)+\sigma\left(\omega,\beta\right)\kappa_z^{(1,z)}\kappa_z^{(2,z)}/i\omega\varepsilon_0\varepsilon_{r\|}^1\varepsilon_{r\|}^2=0$ associated with the $TM_x$, the $TM_y$ or magnetic plasmon, and $TM_z$ modal groups respectively. Moreover, when the two materials surrounding the conducting interface are similar, as for graphene sandwiched by two similar materials the propagation constant for the optical modes is obtained as

$$\beta^x=\omega\varepsilon_0\sqrt{\eta_0^2\varepsilon_{r\|}\mu_r-4\left(\varepsilon_{r\|}/\sigma\left(\omega\right)\right)^2} \tag{1.10a}$$

for the $TM_x$ modes, and

$$\beta^y=\omega\mu_0\sqrt{\eta_0^{-2}\varepsilon_{r\|}\mu_r-\frac{1}{4}\left(\mu_r\sigma\left(\omega\right)\right)^2} \tag{1.10b}$$

For the $TM_y$ modes. Finally for the $TM_z$ modes we have

$$\beta^z=\omega\varepsilon_0\sqrt{\eta_0^2\varepsilon_{r\perp}\mu_r-4\left(\varepsilon_{r\|}/\sigma\left(\omega\right)\right)^2} \tag{1.10c}$$

where we assumed the local approximation for conductivity for the sake of simplicity. For the following calculations, we assume a Drude model for conductivity as $\sigma(\omega)=\left(2e^2 v_{\mathrm{F}}/h\right)\dfrac{\sqrt{\pi n_s}}{1/\tau - i\omega}$, where $\tau$ =500 fs, $v_{\mathrm{F}}$ is the Fermi velocity in graphene, $n_s = 7.37\times10^{12}\mathrm{cm}^{-2}$ is the carrier density. Interestingly, when $\sigma < \eta_0\sqrt{\mu_r/\varepsilon_r}$ , a 2D conducting plane can also carry $TM_y$ plasmons. However, most of the cases $\sigma \ll \eta_0\sqrt{\mu_r/\varepsilon_r}$ , for which $TM_y$ modes are transformed to plane waves in the surrounding medium.

We consider a structure composed of graphene sandwiched by hBN (Fig. 3). The permittivity of hBN is modelled by two Lorentz functions for planar and orthogonal components as described by Woessner et al [30]. $TM_x$ and $TM_z$ modes in such system are nearly degenerate, despite the hyperbolic nature of hBN. These modes sustain giant phase constants, leading to the confinement of graphene plasmons to the area well beyond the diffraction limit. In contrast however, $TM_y$ modes are only loosely bound to the graphene. It should be noted here that for materials without ME, a rather simpler derivation can be considered based on the isofrequency surfaces provided by eq. (1.1) and Maxwell's equations. In this case one directly constructs the solutions at the field level, where the only nonzero components are $E_x$ , $E_z$, and $H_y$. The propagation constant is then obtained as $\beta = \omega\varepsilon_0\sqrt{\eta_0^2\varepsilon_{rzz} - 4\varepsilon_{rzz}\varepsilon_{rxx}/\sigma_x^2}$ (Fig. 4). There is a significant difference between the propagation constants shown in figures 3 and 4. In fact the propagation constant obtained by decomposing the fields are understood by the level repulsion between the modes obtained by decomposing the magnetic vector potential and the planar waves in bulk hBN.

We now assume the optical modes at the interface between hBN and air (Fig. 5). In this system only $TM_x$ modes are propagating, and the $TM_x$ and $TM_z$ modes are no longer degenerate. The optical modes at the interface of a hyperbolic material like hBN and a dielectric are called Dyakonov plasmons. By inserting a graphene layer at the interface, the graphene plasmons are also excited, which poses two individual modes at energies below 160 meV and at energies above 195 eV. The effective wavelength of the graphene plasmon (Dirac plasmons) is much shorter than the wavelength of the light, but it cannot propagate to a long distance, in contrast to the Dyakonov plasmon.

$Bi_2Se_3$ is also an example of a material which is naturally hyperbolic, both at THz frequencies and in the visible range. Moreover, the topological ME effect can also alter the modal dispersion of the optical fields at the interface, though the effect is almost negligible. We use the bulk permittivity components reported by Wu et al. (Fig. 6), and for the surface conductivity we use the same model as for graphene. Dyakonov plasmons at the interface of $Bi_2Se_3$ and air are very much attached to the light cone, which results in less confinement of the optical energy in comparison with hBN. Moreover, only the $A^{xy}$ group supports propagating waves at the interface between $Bi_2Se_3$ and air. Dirac plasmons at the surface of $Bi_2Se_3$ are only excited at energies below 8 meV and above 17 meV, as shown in ref. [27] (Fig. 6). Due to the existence of the topological ME effect, the $y$-component of the electric field is also excited, even when the interface is excited with a $p$-polarized light. This fact leads to the Faraday rotation due to the topological ME effect just at a single interface.

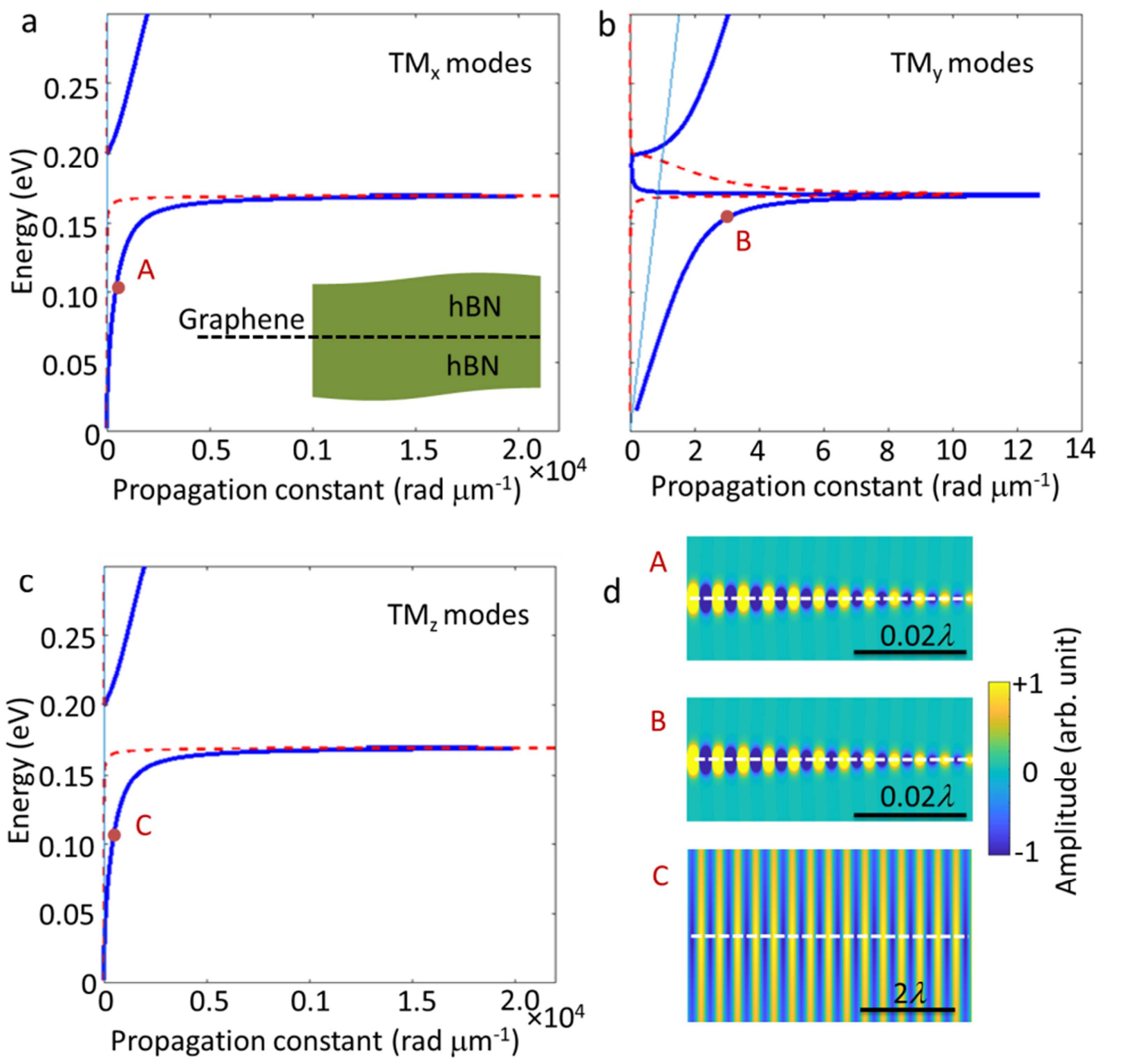

**Fig. 3.** Propogation constant of a graphene sheet sandwiched by hBN for (a) $TM_x$ and (b) $TM_y$, and (c) $TM_z$ modes. $TM_x$ and $TM_z$ modes are nearly degenerate, despite the hyperbolic behaviour of hBN. TMy modes (middle) are loosely bounded to the graphene. Phase constant is shown by the blue solid lines and the attenuation constant by red dashed lines. Optical line in air is shown by cyan solid line. (d) The spatial distribution of the tangential $x$-component of the electric field at depicted energies for each mode.

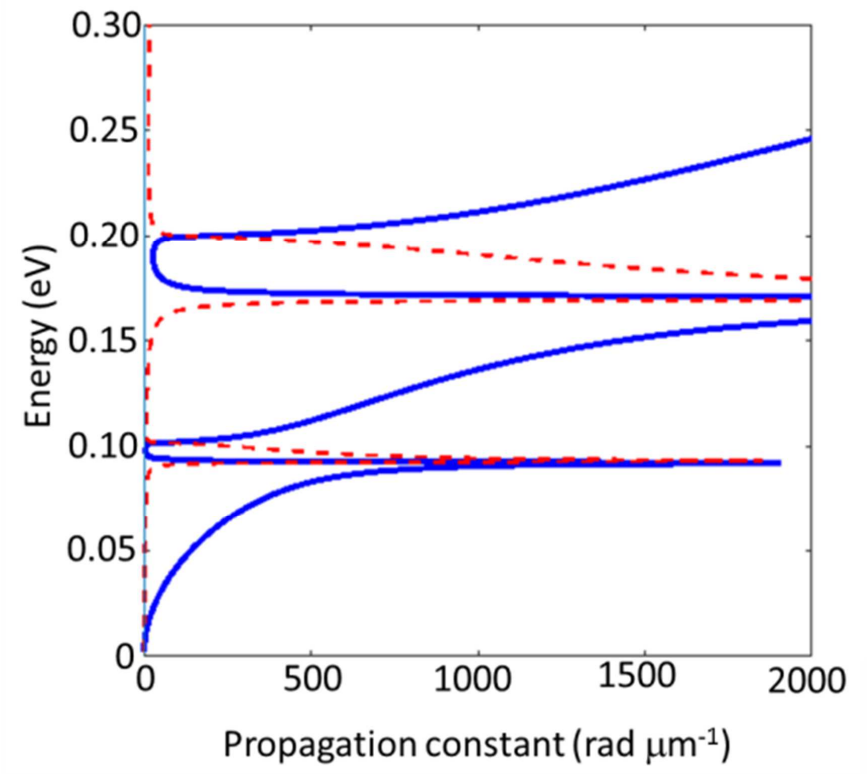

**Fig. 4.** Propagation constant obtained by directly decomposing the fields into the $TM_x$ and $TM_y$ groups, where only $TM_x$ modes lead to propagating waves. The phase and the attenuation constants are shown by blue solid lines and red dashed lines, respectively.

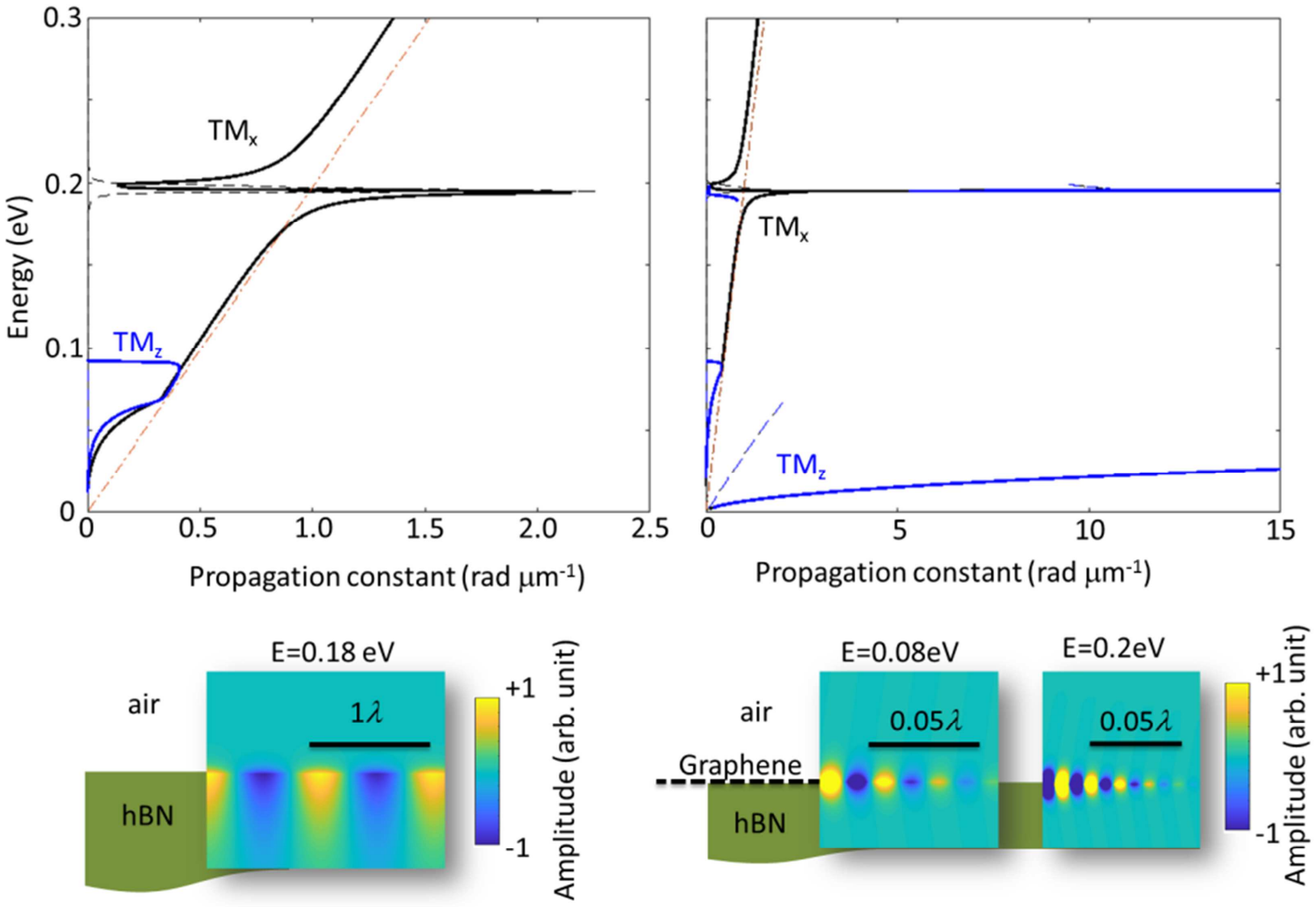

**Fig. 5.** (left) Dispersion of Dyakonov plasmons at the interface of hBN and air. (right) Dispersion of plasmons confined at the hBN/graphene/air structure. Spatial profile of the *x*-component of the electric fields at depicted energies, for the hBN/air interface (left lower panel) and the hBN/graphene/air interface (right lower panel).

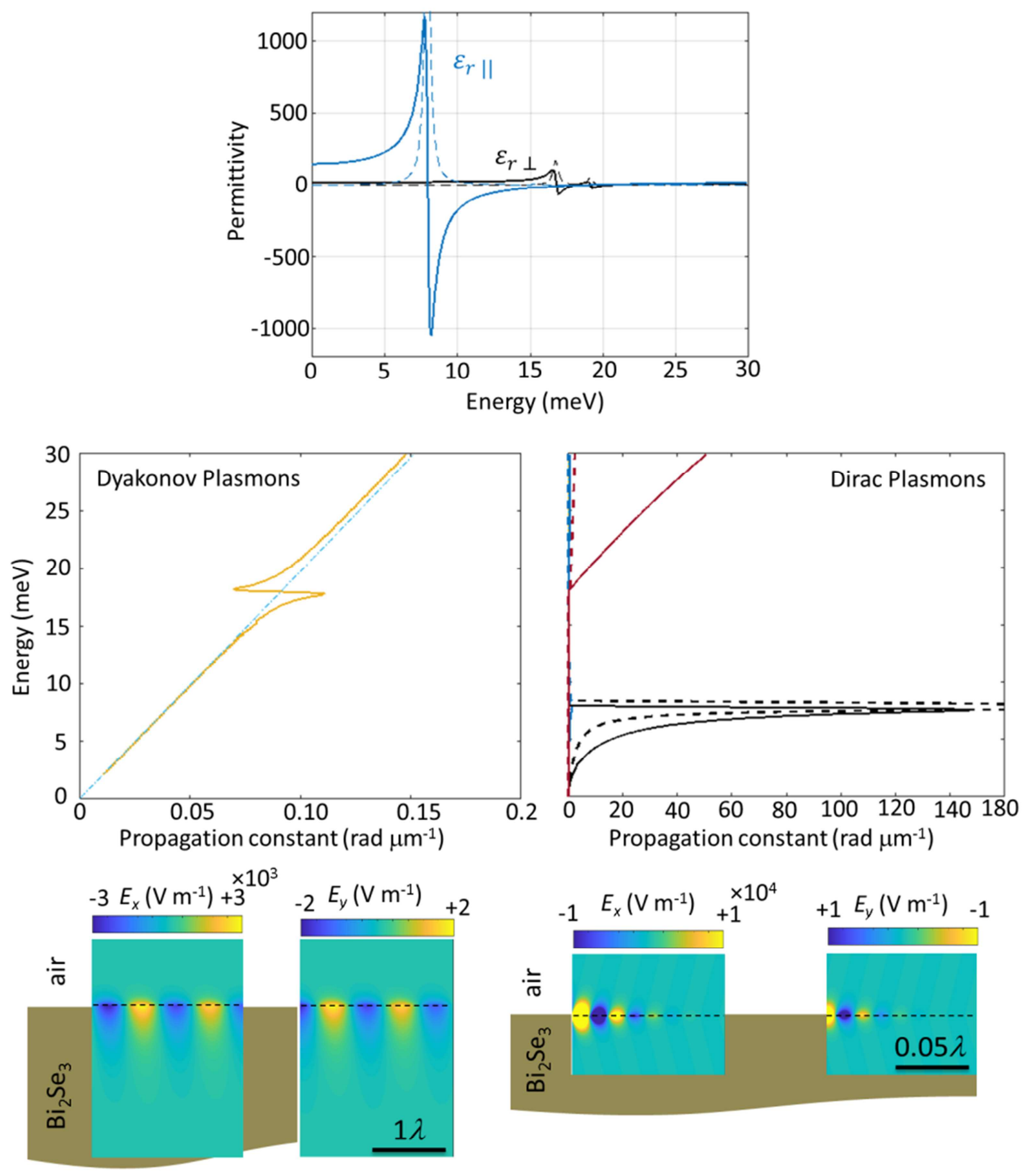

**Fig. 6.** (Top) Permittivity components for $Bi_2Se_3$ in the THz regime. Real parts are shown by solid lines, and the imaginary parts by dashed lines. (Middle left) Dispersion of Dyakonov plasmons at the interface of $Bi_2Se_3$ and air, excluding the Dirac plasmon dispersion. (Bottom left) *x*- and *y*-components of the electric field at the energy of 157 meV. (Middle left) Dispersion of Dirac plasmons at the interface of $Bi_2Se_3$ and air. (Bottom right) *x*- and *y*- components of the electric field at the energy of 22 meV.

Another hyperbolic band for $Bi_2Se_3$ exists at optical frequencies (Fig. 7) [26]. This band is specifically interesting due to the fact that both normal and parallel (to the interface) permittivity components exchange their signs in the visible-frequency range leading to the existence of both type-I and type-II hyperbolic behaviours in the bulk. However, neither Dirac plasmons nor the topological ME effect are excited in this energy range, which is quite above the Fermi energy of the material. The dispersion of Dyakonov plasmons in this energy range is quite attached to the light line, leading to penetration of the evanescent tail of the field at long distances into the air. Although Dyakonov plasmons cannot propagate at long ranges, guiding modes inside a thin film made of hyperbolic materials, and also at the wedges of nanostructures sustain smaller attenuation constants. In the following we investigate the behaviour of optical modes in slab waveguides made of hyperbolic materials with topological ME effect.

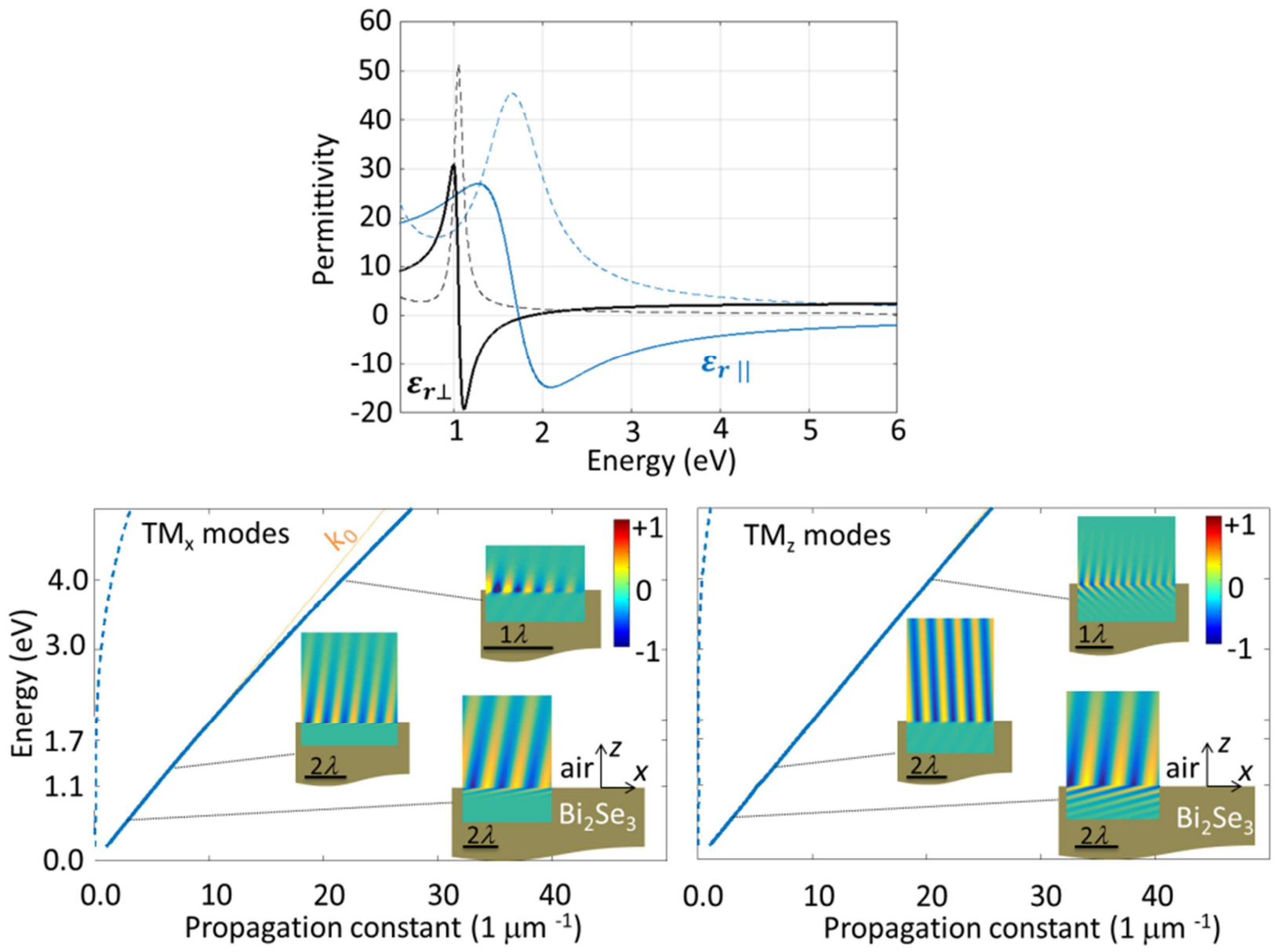

**Fig. 7.** (Top) Permittivity components for $Bi_2Se_3$ at optical ferquencies. Real parts are shown by solid lines, and the imaginary parts by dashed lines. Dispersion of Dyakonov plasmons at the interface of $Bi_2Se_3$ and air for (Bottom left) $TM_x$ modes and (Bottom right) for the $TM_z$ modes. *x*-component of the electric at selected energies is shown at the inset.

## 1.3 Optical modes in a slab waveguide

We consider a slab waveguide spanned within the region $|z| \leq d$ and positioned on top of a substrate (Fig. 8). Two different 2D conducting sheets are positioned at $z = +d$ and $z = -d$, with conductivities $\sigma_u$ and $\sigma_d$, respectively. Here, $u$ and $d$ stand for up and down, respectively. The solution to the propagating wave along the $x$-axis in this heterostructure can be constructed as

$$A_\alpha(\vec{r},\omega)=\begin{cases}\tilde{A}_3^\alpha\, e^{-\kappa_z^a(z-d)}e^{-i\beta x} & \forall z\geq +d\\ \left(\tilde{A}_{2\mathrm{e}}^\alpha\cos\left(\kappa_z^{(2,\alpha)}z\right)+\tilde{A}_{2o}^\alpha\sin\left(\kappa_z^{(2,\alpha)}z\right)\right)e^{-i\beta x} & \forall |z|\leq d\\ \tilde{A}_1^\alpha\, e^{+\kappa_z^s(z+d)}e^{-i\beta x} & \forall z\leq -d\end{cases} \tag{1.11}$$

where $\left(\kappa_z^a\right)^2=\beta^2-\varepsilon_{ra}\mu_{ra}k_0^2$, $\left(\kappa_z^{(2,\alpha)}\right)^2=\varepsilon_{r\alpha\alpha}\mu_r k_0^2-\beta^2$, and $\left(\kappa_z^s\right)^2=\beta^2-\varepsilon_{rs}\mu_{rs}k_0^2$, and $\alpha\in(x,\ y,\ z)$. The possible solutions are again constructed by assuming a pair of potentials as $\left(A_x,A_y\right)$ and $\left(A_y,A_z\right)$ referred to as $A^{xy}$ and $A^{yz}$, respectively. By satisfying the boundary conditions, the characteristic equation for the propagation constant $\beta=\beta'-i\beta''$ is computed. After some straightforward algebraic efforts we derive the following system of equation for $A^{xy}$ group:

$$\begin{bmatrix} c_{11} & c_{12} & c_{13} & c_{14}\\ c_{21} & c_{22} & c_{23} & c_{24}\\ c_{31} & c_{32} & c_{33} & c_{34}\\ c_{41} & c_{42} & c_{43} & c_{44}\end{bmatrix}\begin{bmatrix} A_{2e}^x\\ A_{2o}^x\\ A_{2e}^y\\ A_{2o}^y\end{bmatrix}=0 \tag{1.12}$$

where the matrix elements are given in table 1. Equation (1.12) is a homogeneous system of equations which can lead to nontrivial solutions only if the determinant of the matrix is zero (i.e. when the matrix is singular). Moreover, for $\sigma_u=\sigma_d=0$, the same equations as those reported in ref. [38] are obtained. It is already visible from the symmetry of the system, that the solutions are further decomposed into two subgroups in the case that $\sigma_u=\sigma_d=\sigma$, $\varepsilon_{ra}=\varepsilon_{rs}=\varepsilon_{rd}$, and $\mu_{ra}=\mu_{rs}=\mu_{rd}$. These subgroups for $A^{xy}$ modes are $\left(A_x^{\mathrm{e}},A_y^{\mathrm{o}}\right)$ and $\left(A_x^{\mathrm{o}},A_y^{\mathrm{e}}\right)$ pairs for which the choice of the unknown amplitudes for the vector potential in the region $|z|\leq d$ is represented by $\left(\tilde{A}_{2\mathrm{e}}^x,\tilde{A}_{2\mathrm{o}}^y\right)$ and $\left(\tilde{A}_{2\mathrm{o}}^x,\tilde{A}_{2\mathrm{e}}^y\right)$, respectively. Here the subscripts *e* and *o* stand for even and odd respectively. For $\left(A_x^{\mathrm{e}}, A_y^{\mathrm{o}}\right)$ group, we derive the following characteristic equation

$$\left(\frac{\varepsilon_{rd}}{\varepsilon_{r\parallel}}\frac{\kappa_z^{(x,2)}}{\kappa_z^d}\cot\left(\kappa_z^{(x,2)}d\right)+\frac{1}{\left(\frac{\sigma}{i\omega\varepsilon_0}\frac{\kappa_z^d}{\varepsilon_{rd}}+1\right)}\right)\times \left(1+\frac{\mu_{rd}}{\mu_r}\frac{\kappa_z^{(y,2)}}{\left(i\omega\mu_0\mu_{rd}\sigma+\kappa_z^d\right)}\cot\left(\kappa_z^{(y,2)}d\right)\right)= -\eta_0^2\xi^2\frac{\mu_{rd}}{\varepsilon_{rx}}\frac{1}{\frac{\sigma}{i\omega\varepsilon_0}\frac{\kappa_z^d}{\varepsilon_{rd}}+1}\frac{\kappa_z^{(x,2)}\cot\left(\kappa_z^{(x,2)}d\right)}{i\omega\mu_0\mu_{rd}\sigma+\kappa_z^d} \quad (1.13)$$

where $\left(\kappa_z^d\right)^2=\beta^2-\varepsilon_{rd}\mu_{rd}k_0^2$. For the $\left(A_x^{\mathrm{o}}, A_y^{\mathrm{e}}\right)$ group we have

$$
\begin{aligned}
&\left(\frac{\varepsilon_{rd}}{\varepsilon_{rx}}\frac{\kappa_z^{(x,2)}}{\kappa_z^d}\tan\left(\kappa_z^{(x,2)}d\right)-\frac{1}{\left(\frac{\sigma}{i\omega\varepsilon_0}\frac{\kappa_z^d}{\varepsilon_{rd}}+1\right)}\right)\\
&\times\left(1-\frac{\mu_{rd}}{\mu_r}\frac{\kappa_z^{(y,2)}}{\left(i\omega\,\mu_0\mu_{rd}\sigma+\kappa_z^s\right)}\tan\left(\kappa_z^{(y,2)}d\right)\right)\\
&=-\eta_0^2\,\xi^2\,\frac{\mu_{rd}}{\varepsilon_{rx}}\frac{1}{\frac{\sigma}{i\omega\varepsilon_0}\frac{\kappa_z^d}{\varepsilon_{rd}}+1}\times\frac{\kappa_z^{(x,2)}\tan\left(\kappa_z^{(x,2)}d\right)}{i\omega\,\mu_0\mu_{rd}\sigma+\kappa_z^d}
\end{aligned}
\tag{1.14}
$$

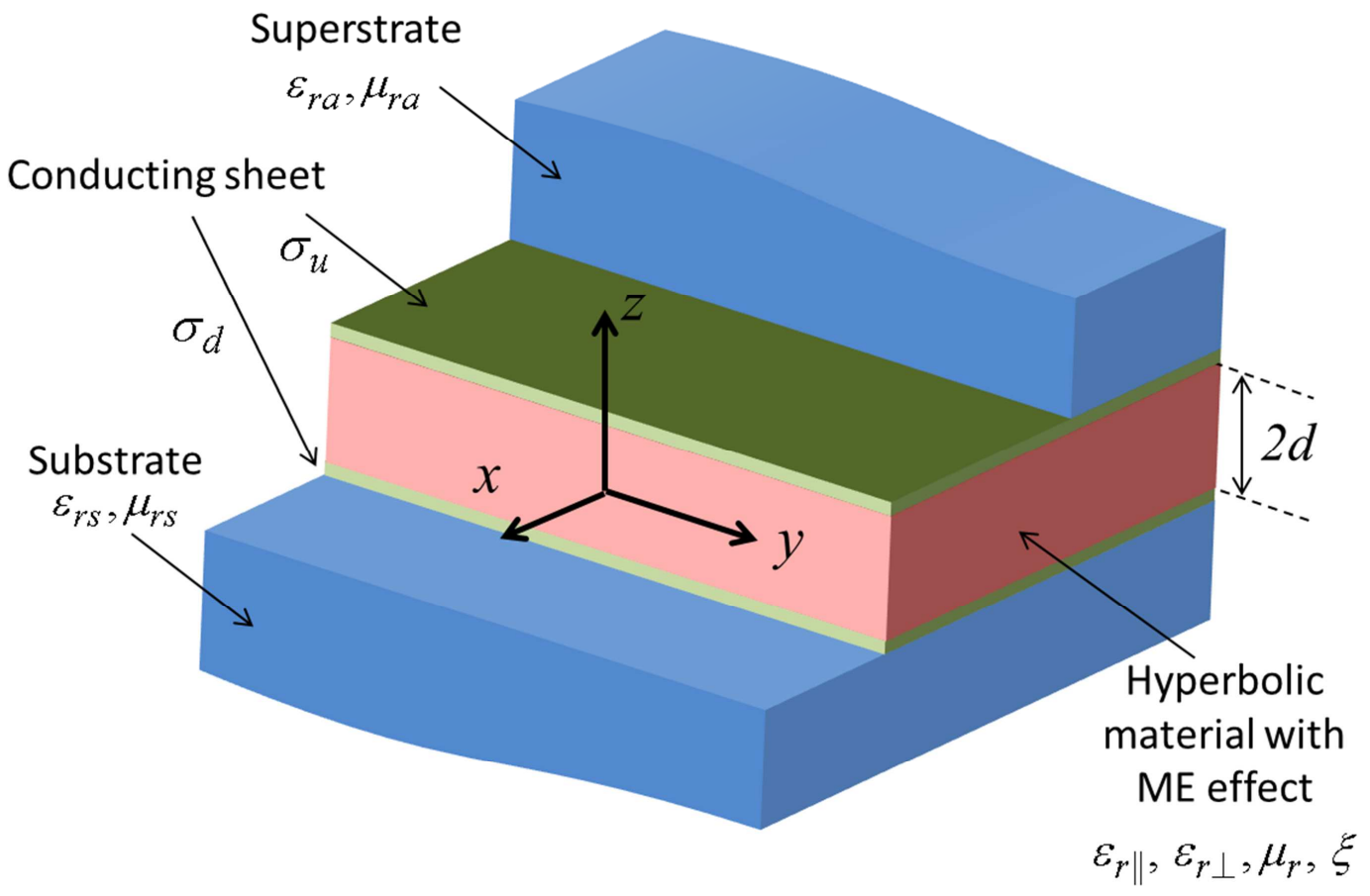

**Fig. 8.** A heterostructure composed of a thin film of hyperbolic material with topological ME effect, sandwiched by 2D conducting sheets positioned at lower and upper surfaces, a superstrate and a substrate.

It is evident from eqs. (1.14) and (1.15), that in the presence of ME effect, we will have a coupling between $x$- and $y$- polarizations. In other words, for $\xi$ =0 we can obtain usual $\mathrm{TM}_x^{\mathrm{e}}$, $\mathrm{TM}_x^{\mathrm{o}}$, $\mathrm{TM}_y^{\mathrm{e}}$, and $\mathrm{TM}_y^{\mathrm{o}}$ modes, with their associated characteristic equations given by

$$\left(\varepsilon_{rd}\kappa_z^{(x,2)}\big/\varepsilon_{r\|}\kappa_z^d\right)\cot\left(\kappa_z^{(x,2)}d\right)=-\left(\left(\sigma\kappa_z^d\big/i\omega\varepsilon_0\varepsilon_{rd}\right)+1\right)^{-1} \tag{1.15a}$$

$$\left(\varepsilon_{rd}\kappa_z^{(x,2)}\big/\varepsilon_{rx}\kappa_z^d\right)\tan\left(\kappa_z^{(x,2)}d\right)=\left(\left(\sigma\kappa_z^d\big/i\omega\varepsilon_0\varepsilon_{rd}\right)+1\right)^{-1} \tag{1.15b}$$

$$\mu_{rd}\kappa_z^{(y,2)}\tan\left(\kappa_z^{(y,2)}d\right)=\mu_r\left(i\omega\,\mu_0\mu_{rd}\sigma+\kappa_z^s\right) \tag{1.15c}$$

and

$$\mu_{rd}\kappa_z^{(y,2)}\cot\left(\kappa_z^{(y,2)}d\right)=-\mu_r\left(i\omega\,\mu_0\mu_{rd}\sigma+\kappa_z^d\right) \tag{1.15d}$$

respectively.

For the $A^{yz}$ modes, the following homogeneous system of equations is obtained:

**Table 1.** Matrix elements for eq. (1.12). Here $\kappa_z^{a,x}=\kappa_z^a\left[1+\left(\sigma_u\kappa_z^a/i\omega\varepsilon_0\varepsilon_{ra}\right)\right]$, $\kappa_z^{s,x}=\kappa_z^s\left[1+\left(\sigma_d\kappa_z^s/i\omega\varepsilon_0\varepsilon_{rs}\right)\right]$, $\kappa_z^{a,y}=\kappa_z^a+i\omega\mu_0\mu_{ra}\sigma_u$, $\kappa_z^{s,y}=\kappa_z^s+i\omega\mu_0\mu_{rs}\sigma_d$.

| $c_{ij}$ | 1 | 2 | 3 | 4 |
|---|---|---|---|---|
| 1 | $-\frac{\varepsilon_{ra}\mu_{ra}}{\varepsilon_{r\parallel}\mu_r}\left(\frac{\kappa_z^{(x,2)}}{\kappa_z^a}\right)^2 \times\cos\left(\kappa_z^{(x,2)}d\right)+$ $-\frac{\mu_{ra}}{\mu_r}\frac{\kappa_z^{(x,2)}}{\kappa_z^{a,x}} \times\sin\left(\kappa_z^{(x,2)}d\right)$ | $-\frac{\varepsilon_{ra}\mu_{ra}}{\varepsilon_{r\parallel}\mu_r}\left(\frac{\kappa_z^{(x,2)}}{\kappa_z^a}\right)^2 \times\sin\left(\kappa_z^{(x,2)}d\right)+$ $+\frac{\mu_{ra}}{\mu_r}\frac{\kappa_z^{(x,2)}}{\kappa_z^{a,x}} \times\cos\left(\kappa_z^{(x,2)}d\right)$ | $\frac{i\omega\mu_0\mu_{ra}\xi}{\kappa_z^{a,x}} \times\cos\left(\kappa_z^{(y,2)}d\right)$ | $\frac{i\omega\mu_0\mu_{ra}\xi}{\kappa_z^{a,x}} \times\sin\left(\kappa_z^{(y,2)}d\right)$ |
| 2 | $-\frac{\varepsilon_{rs}\mu_{rs}}{\varepsilon_{r\parallel}\mu_r}\left(\frac{\kappa_z^{(x,2)}}{\kappa_z^s}\right)^2 \times\cos\left(\kappa_z^{(x,2)}d\right)$ $-\frac{\mu_{rs}}{\mu_r}\frac{\kappa_z^{(x,2)}}{\kappa_z^{s,x}} \times\sin\left(\kappa_z^{(x,2)}d\right)$ | $-\frac{\varepsilon_{rs}\mu_{rs}}{\varepsilon_{r\parallel}\mu_r}\left(\frac{\kappa_z^{(x,2)}}{\kappa_z^s}\right)^2 \times\sin\left(\kappa_z^{(x,2)}d\right)$ $+\frac{\mu_{rs}}{\mu_r}\frac{\kappa_z^{(x,2)}}{\kappa_z^{s,x}} \times\cos\left(\kappa_z^{(x,2)}d\right)$ | $\frac{-i\omega\mu_0\mu_{rs}\xi}{\kappa_z^{s,x}} \times\cos\left(\kappa_z^{(y,2)}d\right)$ | $\frac{i\omega\mu_0\mu_{rs}\xi}{\kappa_z^{s,x}} \times\sin\left(\kappa_z^{(y,2)}d\right)$ |
| 3 | $\frac{\xi}{\kappa_z^{a,y}}\frac{\left(\kappa_z^{(x,2)}\right)^2}{i\omega\varepsilon_0\varepsilon_{r\parallel}}\frac{\mu_{ra}}{\mu_r} \times\cos\left(\kappa_z^{(x,2)}d\right)$ | $\frac{\xi}{\kappa_z^{a,y}}\frac{\left(\kappa_z^{(x,2)}\right)^2}{i\omega\varepsilon_0\varepsilon_{r\parallel}}\frac{\mu_{ra}}{\mu_r} \times\sin\left(\kappa_z^{(x,2)}d\right)$ | $\cos\left(\kappa_z^{(y,2)}d\right)$ $-\frac{\kappa_z^{(y,2)}}{\kappa_z^{a,y}}\frac{\mu_{ra}}{\mu_r} \times\sin\left(\kappa_z^{(y,2)}d\right)$ | $\sin\left(\kappa_z^{(y,2)}d\right)$ $+\frac{\kappa_z^{(y,2)}}{\kappa_z^{a,y}}\frac{\mu_{ra}}{\mu_r} \times\cos\left(\kappa_z^{(y,2)}d\right)$ |
| 4 | $\frac{-\xi}{\kappa_z^{s,y}}\frac{\left(\kappa_z^{(x,2)}\right)^2}{i\omega\varepsilon_0\varepsilon_{r\parallel}}\frac{\mu_{rs}}{\mu_r} \times\cos\left(\kappa_z^{(x,2)}d\right)$ | $\frac{+\xi}{\kappa_z^{s,y}}\frac{\left(\kappa_z^{(x,2)}\right)^2}{i\omega\varepsilon_0\varepsilon_{r\parallel}}\frac{\mu_{rs}}{\mu_r} \times\sin\left(\kappa_z^{(x,2)}d\right)$ | $\cos\left(\kappa_z^{(y,2)}d\right)$ $-\frac{\kappa_z^{(y,2)}}{\kappa_z^{s,y}}\frac{\mu_{rs}}{\mu_r}\times$ $\sin\left(\kappa_z^{(y,2)}d\right)$ | $\sin\left(\kappa_z^{(y,2)}d\right)$ $+\frac{\kappa_z^{(y,2)}}{\kappa_z^{s,y}}\frac{\mu_{rs}}{\mu_r}\times$ $\cos\left(\kappa_z^{(y,2)}d\right)$ |

$$\begin{bmatrix} c_{11} & c_{12} & c_{13} & c_{14} \\ c_{21} & c_{22} & c_{23} & c_{24} \\ c_{31} & c_{32} & c_{33} & c_{34} \\ c_{41} & c_{42} & c_{43} & c_{44} \end{bmatrix} \begin{bmatrix} A_{2e}^{z} \\ A_{2o}^{z} \\ A_{2e}^{y} \\ A_{2o}^{y} \end{bmatrix} = 0 \tag{1.16}$$

with the matrix elements provided in table 2. Obviously we should solve for the zeros of the determinant of the matrix in (1.16) to obtain the propagation constant. As for $A^{xy}$ modes, $A^{yz}$ modes can be also further decomposed into two subgroups for a symmetric system; i.e. are $\left(\tilde{A}_{2\mathrm{e}}^{y}, \tilde{A}_{2\mathrm{e}}^{z}\right)$ and $\left(\tilde{A}_{2\mathrm{o}}^{y}, \tilde{A}_{2\mathrm{o}}^{z}\right)$ pairs. The characteristic equation for the propagation constant associated with $\left(\tilde{A}_{2\mathrm{e}}^{y}, \tilde{A}_{2\mathrm{e}}^{z}\right)$ can be obtained as

$$\begin{aligned} &\left[ \frac{\varepsilon_{rd}\kappa_z^{(z,2)}}{\varepsilon_{r\parallel}\kappa_z^{d}} - \frac{1}{\left(\dfrac{\sigma}{i\omega\varepsilon_0}\dfrac{\kappa_z^{d}}{\varepsilon_{rd}}+1\right)}\cot\left(\kappa_z^{(z,2)}d\right)\right] \\ &\times\left[1-\frac{\mu_{rd}}{\mu_r}\frac{\kappa_z^{(y,2)}}{\left(i\omega\mu_0\mu_{rd}\sigma+\kappa_z^{d}\right)}\tan\left(\kappa_z^{(y,2)}d\right)\right] = \\ &-\eta_0^2\xi^2\frac{\mu_{rd}}{\varepsilon_{r\parallel}}\frac{1}{\left(\dfrac{\sigma}{i\omega\varepsilon_0}\dfrac{\kappa_z^{d}}{\varepsilon_{rd}}+1\right)}\frac{\kappa_z^{(z,2)}}{\left(i\omega\mu_0\mu_{rd}\sigma+\kappa_z^{d}\right)} \end{aligned} \tag{1.17}$$

For $\left(\tilde{A}_{2\mathrm{o}}^{y}, \tilde{A}_{2\mathrm{o}}^{z}\right)$ modes, the characteristic equation is obtained as

$$\begin{aligned} &\left[ \frac{\varepsilon_{rd}\kappa_z^{(z,2)}}{\varepsilon_{r\parallel}\kappa_z^{d}} + \frac{1}{\left(\dfrac{\sigma}{i\omega\varepsilon_0}\dfrac{\kappa_z^{d}}{\varepsilon_{rd}}+1\right)}\tan\left(\kappa_z^{(z,2)}d\right)\right] \\ &\times\left[1+\frac{\mu_{rd}}{\mu_r}\frac{\kappa_z^{(y,2)}}{\left(i\omega\mu_0\mu_{rd}\sigma+\kappa_z^{d}\right)}\cot\left(\kappa_z^{(y,2)}d\right)\right] = \\ &-\eta_0^2\xi^2\frac{\mu_{rd}}{\varepsilon_{r\parallel}}\frac{1}{\left(\dfrac{\sigma}{i\omega\varepsilon_0}\dfrac{\kappa_z^{d}}{\varepsilon_{rd}}+1\right)}\frac{\kappa_z^{(z,2)}}{\left(i\omega\mu_0\mu_{rd}\sigma+\kappa_z^{d}\right)} \end{aligned} \tag{1.18}$$

**Table 2.** Matrix elements for eq. (1.12). Here $\kappa_z^{a,x} = 1 + \left(\sigma_u \kappa_z^a / i\omega\varepsilon_0\varepsilon_{ra}\right)$, $\kappa_z^{s,x} = 1 + \left(\sigma_d \kappa_z^s / i\omega\varepsilon_0\varepsilon_{rs}\right)$, $\kappa_z^{a,y} = \kappa_z^a + i\omega\,\mu_0\mu_{ra}\sigma_u$, $\kappa_z^{s,y} = \kappa_z^s + i\omega\,\mu_0\mu_{rs}\sigma_d$.

| $c_{ij}$ | $j$=1 | $j$=2 | $j$=3 | $j$=4 |
|---|---|---|---|---|
| $i$=1 | $\frac{\varepsilon_{ra}\mu_{ra}}{\varepsilon_{rx}\mu_r}\frac{\kappa_z^{(z,2)}}{\kappa_z^a}\times\sin\left(\kappa_z^{(z,2)}d\right)-\frac{1}{\mu_r}\frac{\mu_{ra}}{\kappa_z^{a,x}}\times\cos\left(\kappa_z^{(z,2)}d\right)$ | $-\frac{\varepsilon_{ra}\mu_{ra}}{\varepsilon_{rx}\mu_r}\frac{\kappa_z^{(z,2)}}{\kappa_z^a}\times\cos\left(\kappa_z^{(z,2)}d\right)-\frac{1}{\mu_r}\frac{\mu_{ra}}{\kappa_z^{a,x}}\times\sin\left(\kappa_z^{(z,2)}d\right)$ | $-\frac{\omega\mu_0}{\beta}\frac{\mu_{ra}\xi}{\kappa_z^{a,x}}\times\cos\left(\kappa_z^{(y,2)}d\right)$ | $-\frac{\omega\mu_0}{\beta}\frac{\mu_{ra}\xi}{\kappa_z^{a,x}}\times\sin\left(\kappa_z^{(y,2)}d\right)$ |
| $i$=2 | $-\frac{\varepsilon_{rs}\mu_{rs}}{\varepsilon_{rx}\mu_r}\frac{\kappa_z^{(z,2)}}{\kappa_z^s}\times\sin\left(\kappa_z^{(z,2)}d\right)+\frac{1}{\mu_r}\frac{\mu_{rs}}{\kappa_z^{s,x}}\times\cos\left(\kappa_z^{(z,2)}d\right)$ | $-\frac{\varepsilon_{rs}\mu_{rs}}{\varepsilon_{rx}\mu_r}\frac{\kappa_z^{(z,2)}}{\kappa_z^s}\times\cos\left(\kappa_z^{(z,2)}d\right)-\frac{1}{\mu_r}\frac{\mu_{rs}}{\kappa_z^{s,x}}\times\sin\left(\kappa_z^{(z,2)}d\right)$ | $+\frac{\omega\mu_0}{\beta}\frac{\mu_{rs}\xi}{\kappa_z^{s,x}}\times\cos\left(\kappa_z^{(y,2)}d\right)$ | $-\frac{\omega\mu_0}{\beta}\frac{\mu_{rs}\xi}{\kappa_z^{s,x}}\times\sin\left(\kappa_z^{(y,2)}d\right)$ |
| $i$=3 | $+\xi\frac{\beta}{\omega\varepsilon_0}\frac{\mu_{ra}}{\varepsilon_{r\parallel}\mu_r}\frac{\kappa_z^{(z,2)}}{\kappa_z^{a,y}}\times\sin\left(\kappa_z^{(z,2)}d\right)$ | $-\xi\frac{\beta}{\omega\varepsilon_0}\frac{\mu_{ra}}{\varepsilon_{r\parallel}\mu_r}\frac{\kappa_z^{(z,2)}}{\kappa_z^{a,y}}\times\cos\left(\kappa_z^{(z,2)}d\right)$ | $\cos\left(\kappa_z^{(y,2)}d\right)-\frac{\mu_{ra}}{\mu_r}\frac{\kappa_z^{(y,2)}}{\kappa_z^{a,y}}\times\sin\left(\kappa_z^{(y,2)}d\right)$ | $\sin\left(\kappa_z^{(y,2)}d\right)+\frac{\mu_{ra}}{\mu_r}\frac{\kappa_z^{(y,2)}}{\kappa_z^{a,y}}\times\cos\left(\kappa_z^{(y,2)}d\right)$ |
| $i$=4 | $-\xi\frac{\beta}{\omega\varepsilon_0}\frac{\mu_{rs}}{\varepsilon_{r\parallel}\mu_r}\frac{\kappa_z^{(z,2)}}{\kappa_z^{s,y}}\times\sin\left(\kappa_z^{(z,2)}d\right)$ | $+\xi\frac{\beta}{\omega\varepsilon_0}\frac{\mu_{rs}}{\varepsilon_{r\parallel}\mu_r}\frac{\kappa_z^{(z,2)}}{\kappa_z^{s,y}}\times\cos\left(\kappa_z^{(z,2)}d\right)$ | $-\cos\left(\kappa_z^{(y,2)}d\right)+\frac{1}{\mu_r}\frac{\mu_{rs}\kappa_z^{(y,2)}}{\kappa_z^{s,y}}\times\sin\left(\kappa_z^{(y,2)}d\right)$ | $+\sin\left(\kappa_z^{(y,2)}d\right)+\frac{\mu_{rs}}{\mu_r}\frac{\kappa_z^{(y,2)}}{\kappa_z^{s,y}}\times\cos\left(\kappa_z^{(y,2)}d\right)$ |

We will obtain here the solutions for the $A^{xy}$ and $A^{yz}$ optical modes in a heterostructure composed of $SiO_2$/graphene/hBN/graphene/air (Fig. 9). Due to the lack of the ME effect, we can decompose the solutions into $TM_x$, $TM_y$, and $TM_z$ modes, where only $TM_x$ and $TM_z$ modes lead to evanescent waves. We consider a thickness of 20 nm for the hBN, which allows for the hybridization of the Dirac plasmons supported by graphene layers. Interestingly, the lower Dirac plasmon band is hybridized into three distinguished modes (compare Fig. 9 with Fig. 5), where the first, third, and fourth modes are quasi-symmetric, in contrast to the second mode which is quasi-antisymmetric. This is understood from the computed field profiles shown in panels A, B, C, and D. By symmetric and antisymmetric modes we mean here the spatial symmetry of the vector potential. Interestingly, the presence of the substrate only slightly affects the dispersion and spatial profile of the Dirac plasmons. Moreover, due to the thickness of the hBN, it is only the quasi- symmetric hyperbolic mode which is excited. The dielectric function of $SiO_2$ is dispersive at THz frequencies, which affects the hyperbolic plasmons at the hBN/$SiO_2$ interface. Moreover, It is only in the energy range of 0.135 eV< $E$ <0.148 eV, that the interface plasmons are bound to the hBN thin film. In contrast to the hyperbolic plasmons (Fig. 9, upper right panel), $A^{xy}$ and $A^{yz}$ Dirac plasmons are degenerate modes (Fig. 9, upper left panel).

Finally, we consider the most complex case, where the material supports surface states as well as the topological ME effect. Thin films composed of topological insulators are such examples. Here we compute the $A^{xy}$ and $A^{yz}$ modes for a thin $Bi_2Se_3$ film ($d$ = 10 nm) positioned on a glass substrate, within the frequency range of 2 meV to 40 meV and the phase constants up to 200 rad μm$^{-1}$ (Fig. 10). Since the ME effect is present here, $TM_x$ and $TM_y$ modes, as well as $TM_y$ and $TM_z$ modes are coupled with each other and lead to the formation of new groups, with their characteristic equations given by eqs. (1.12) and (1.15), respectively. In contrast to hBN, however, the lower Dirac plasmon dispersion band does not demonstrate any hybridization into symmetric and antisymmetric modes. The lower Dirac plasmon band is a quasi $\left(\tilde{A}_{2\mathrm{o}}^{x}, \tilde{A}_{2\mathrm{e}}^{y}\right)$ mode, which is understood from the computed field profile at the energy–momentum point marked by A. However, the upper Dirac plasmon band is hybridized into two quasi $\left(\tilde{A}_{2\mathrm{o}}^{x}, \tilde{A}_{2\mathrm{e}}^{y}\right)$ and quasi $\left(\tilde{A}_{2e}^{x}, \tilde{A}_{2o}^{y}\right)$ modes (Fig. 10 Panels B and C).

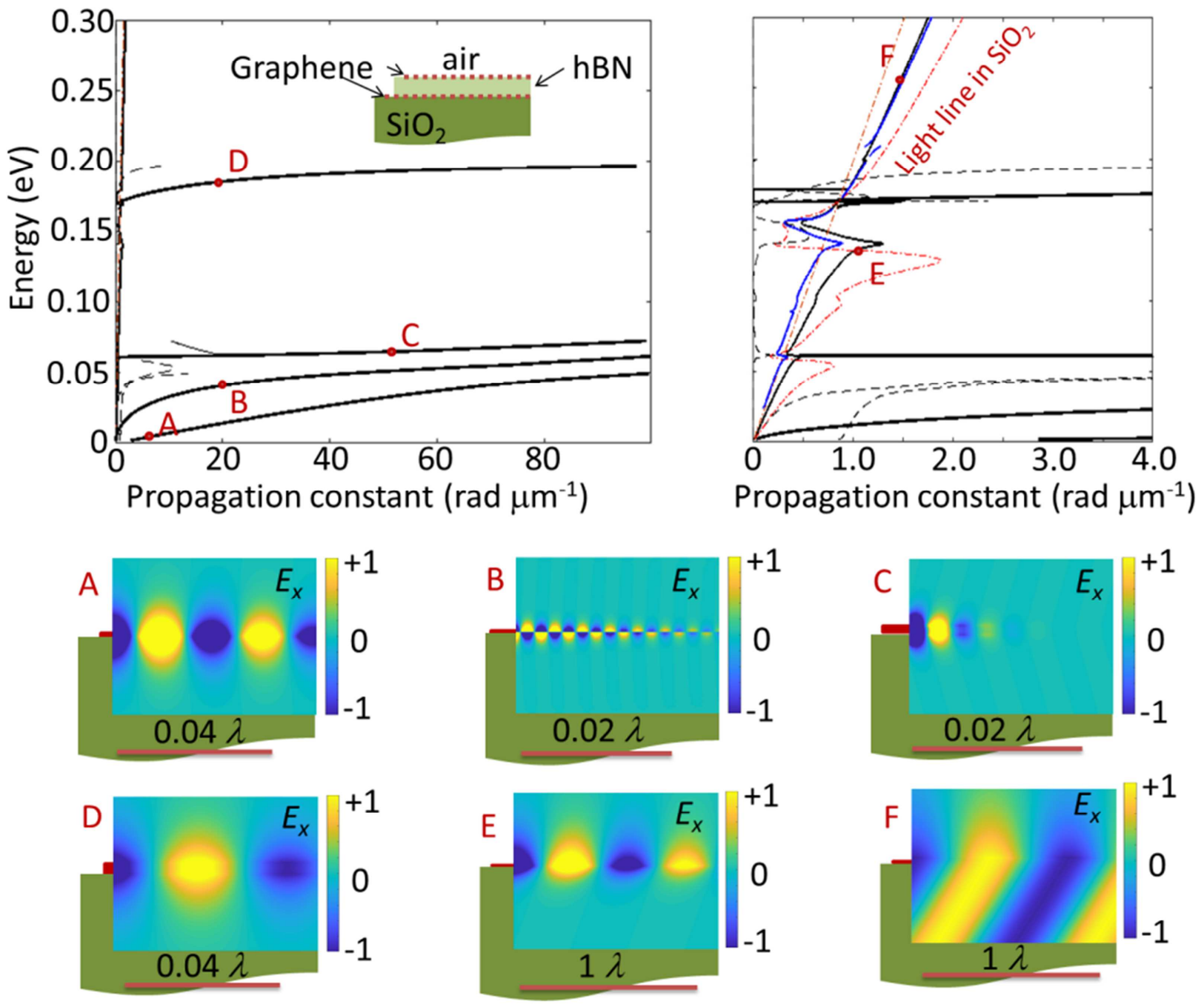

**Fig. 9.** $A^{xy}$ and $A^{yz}$ modes in a heterostructure composed of $SiO_2$/graphene/hBN/air, as shown in the inset. (top left) computed propagation constant for hyperbolic and Dirac plasmons. (top right) Zoom into the lower propagation constant regime to show hyperbolic modes. Phase constant is shown by the solid lines and attenuation constant with dashed lines. Light lines in free space and $SiO_2$ are shown by dashed-dotted lines. (Bottom) spatial profile of the *x*-component of the electric field at the marked energy–momentum points in the phase diagram. The dielectric function for $SiO_2$ is adapted from ref. [39]. $A^{xy}$ and $A^{yz}$ modes in the top right panel are shown by black and blue lines, respectively. $A^{xy}$ and $A^{yz}$ Dirac plasmon modes are degenerate.

The topological ME index given by $\xi = \alpha\theta/\eta_0\pi$ is only of the order of $10^{-6}$, which leads to a small Faraday rotation of the order of 0.0073 rad [32], at a single interface. It was conjectured elsewhere that due to the presence of a second interface which leads to a negative rotation, the overall Faraday rotation might be not because of the topological ME effect, but because of the quantum Faraday effect [40]. However, both symmetric and antisymmetric optical modes can be excited in a thin film composed of the topological insulators. Depending on the coupling of symmetric or antisymmetric modes to the excitation field in the THz spectroscopy apparatus, the situation might lead to a net rotation of the polarization, or the cancellation of the rotations at both interfaces, respectively.

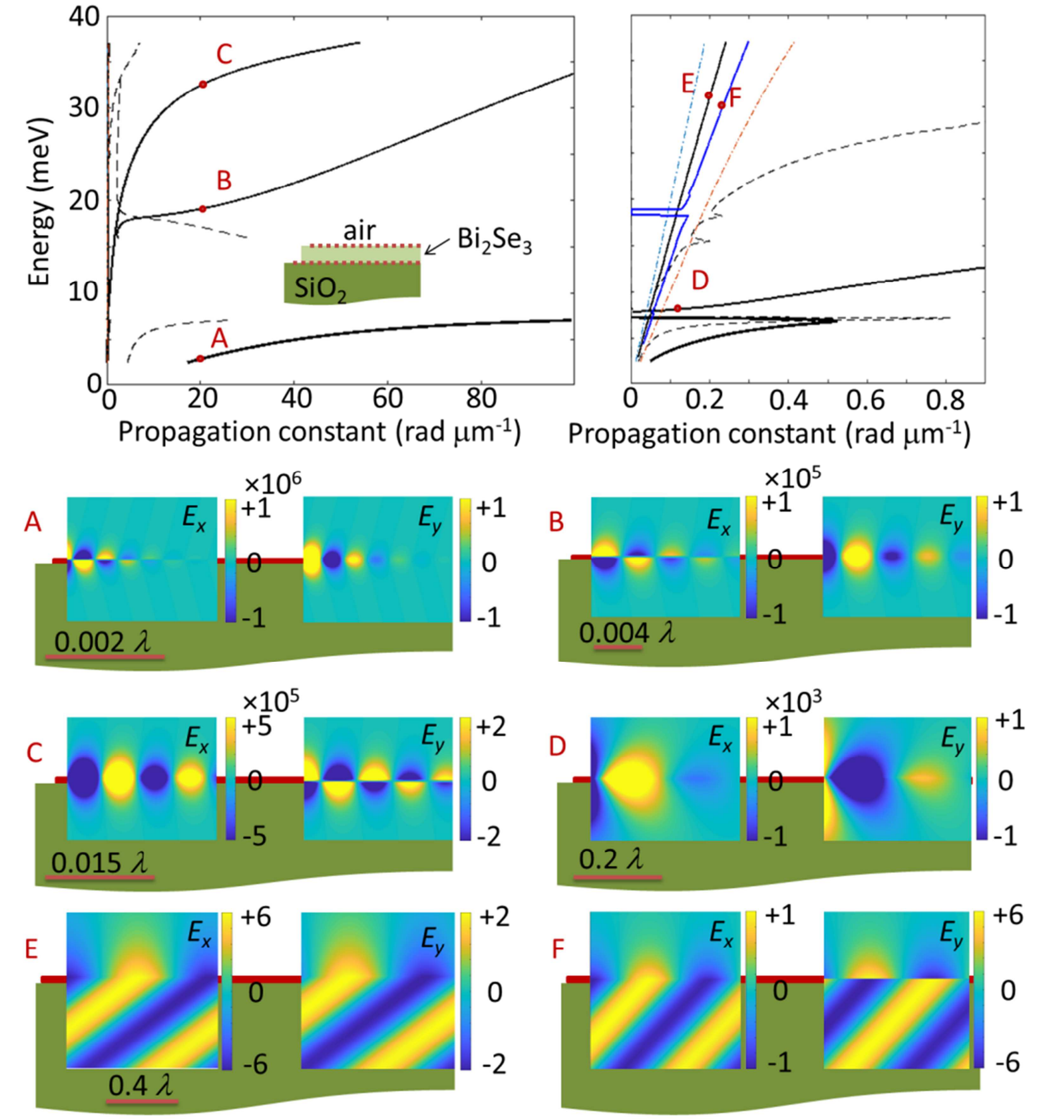

**Fig. 10.** $A^{xy}$ and $A^{yz}$ modes in a heterostructure composed of $SiO_2/Bi_2Se_3$/air, as shown in the inset. (top left) computed propagation constant for hyperbolic and Dirac plasmons including phase constants up to $10^8$ rad μm. (top right) Zoom into the lower propagation constant regime to show hyperbolic modes. The phase constant is shown by the solid lines and attenuation constant with dashed lines. Light lines in free space and $SiO_2$ are shown by dashed-dotted lines. (Bottom) Spatial profile of the *x*- and *y*-components of the electric field at the marked energy–momentum points in the phase diagram. The dielectric function for $SiO_2$ is adapted from ref. [39]. $A^{xy}$ and $A^{yz}$ modes in the top right panel are shown by black and blue lines, respectively. $A^{xy}$ and $A^{yz}$ Dirac plasmons modes are degenerate.

In order to better show the effect of the ME effect on the optical modes supported by our considered system, we computed the spatial profile of both *x*- and *y*- components of the electric field. In fact in the absence of the ME effect $E_y = 0$. Interestingly, the ME effect only slightly affects Dirac plasmon modes, as well as optical waves at higher momentum, as understood by comparing the magnitude of the excited $E_x$ and $E_y$ field components. However, hyperbolic plasmons are greatly affected by the ME effect in such a way that the excited $E_x$ and $E_y$ field components are of the same order of magnitude.

## 1.4 Summary and Outlook

As a summary, we computed in this chapter the dispersion of optical modes excited at the interface and thin films, composed of hyperbolic materials exhibiting a topological ME effect. Several examples were discussed covering topological insulators and naturally hyperbolic materials. We considered a novel grouping of the optical modes, which allows us to directly construct the solutions for the wave equations in a medium with ME effect, for which we allowed for a generalization of the gauge theory to further consider the chiral index of the material. However, in the absence of the chiral ME effect the generalized gauge theory is simplified to the widely accepted Lorentz gauge. The proposed methods and the investigations here clearly demonstrate the ultrahigh momentum modes supported by graphene and surface states in topological insulators, which can provide a platform for novel photonic circuitries including Dirac plasmons, like ultrasmall high-quality resonators and waveguides.

**Acknowledgments**

The author gratefully acknowledges the support from the Stuttgart Center for Electron Microscopy, especially Wilfried Sigle and Peter A. van Aken for fruitful discussions.